\documentclass[runningheads]{llncs}
\usepackage[T1]{fontenc}
\usepackage{graphicx}
\usepackage[table,dvipsnames,svgnames]{xcolor}
\usepackage{tikz}
\usepackage{wrapfig}
\usepackage[inline]{enumitem}
\usepackage{subcaption}   %
\usepackage{booktabs}
\usepackage{siunitx}
\usepackage[colorinlistoftodos,textsize=small]{todonotes}

\usetikzlibrary{arrows.meta,positioning,backgrounds,fit}
\usepackage{hyperref}
\usepackage{color}

\usepackage{algorithm}
\usepackage{algorithmic}

\usepackage{booktabs}   
\usepackage{multirow} 
\usepackage{amsmath}   
\usepackage{amssymb}

\usepackage{siunitx}       %
\usepackage{array}
\usepackage{adjustbox}
\usepackage{colortbl}

\usepackage{threeparttable}
\usepackage{makecell}
\usepackage{textcomp}   %
\usepackage{pifont}     %

\usepackage{newfloat}
\usepackage{listings}

\usepackage{enumitem}

\usepackage{cite}
\usepackage[numbers,sort]{natbib}

\newcommand{\miniskip}{\vspace*{-.5\baselineskip}}
\newcommand{\shrink}{\vspace*{-.9\baselineskip}}

\begin{document}

\definecolor{PS1}{HTML}{fb9a99}
\definecolor{PS2}{HTML}{a6cee3}
\definecolor{PS3}{HTML}{b2df8a}

\newcommand{\ourparagraph}[1]{\smallskip\noindent\emph{#1}}

\title{OrLog: Resolving Complex Queries with LLMs and Probabilistic Reasoning}

\titlerunning{OrLog}

\author{
Mohanna Hoveyda\inst{1}\orcidID{0009-0003-8027-6575} \and
Jelle Piepenbrock\inst{2}\orcidID{0000-0002-8385-9157} \and
Arjen P.\ de Vries\inst{1}\orcidID{0000-0002-2888-4202} \and
Maarten de Rijke\inst{3}\orcidID{0000-0002-1086-0202} \and
Faegheh Hasibi\inst{1}\orcidID{0009-0006-9986-482X}
}

\authorrunning{Hoveyda et al.}

\institute{
Radboud University, Nijmegen, The Netherlands\\
\email{\{mohanna.hoveyda,arjen.devries,faegheh.hasibi\}@ru.nl}
\and
Eindhoven University of Technology, Eindhoven, The Netherlands\\
\email{j.h.piepenbrock@tue.nl}
\and
University of Amsterdam, Amsterdam, The Netherlands\\
\email{m.derijke@uva.nl}
}

\maketitle              

\begin{abstract}
Resolving complex information needs that come with multiple con\-straints should consider enforcing the logical operators encoded in the query (i.e., conjunction, disjunction, negation) on the candidate answer set. Current retrieval systems either ignore these constraints in neural embeddings or approximate them in a generative reasoning process that can be inconsistent and unreliable. 
Although well-suited to structured reasoning, existing neuro‑symbolic approaches remain confined to formal logic or mathematics problems as they often assume unambiguous queries and access to complete evidence, conditions rarely met in information retrieval. 
To bridge this gap, we introduce \textbf{OrLog}, a neuro-symbolic retrieval framework that decouples 
predicate-level plausibility estimation from logical reasoning: %
a large language model (LLM) provides plausibility scores for atomic predicates in one decoding-free forward pass, from which a probabilistic reasoning engine derives the posterior probability of query satisfaction. 
We evaluate OrLog across multiple backbone LLMs, varying levels of access to external knowledge, and a range of logical constraints, and compare it against base retrievers and LLM-as-reasoner methods.
Provided with entity descriptions, OrLog can significantly boost top‑rank precision compared to LLM reasoning with larger gains
on disjunctive queries. OrLog is also more efficient, cutting mean tokens by $\sim$90\% per query–entity pair. These results demonstrate that generation‑free predicate plausibility estimation combined with probabilistic reasoning enables constraint‑aware retrieval that outperforms monolithic reasoning while using far fewer tokens.

\end{abstract}

\begin{figure*}[ht]
  \centering
  \includegraphics[trim=0 0 0 0, clip, width=1\textwidth]{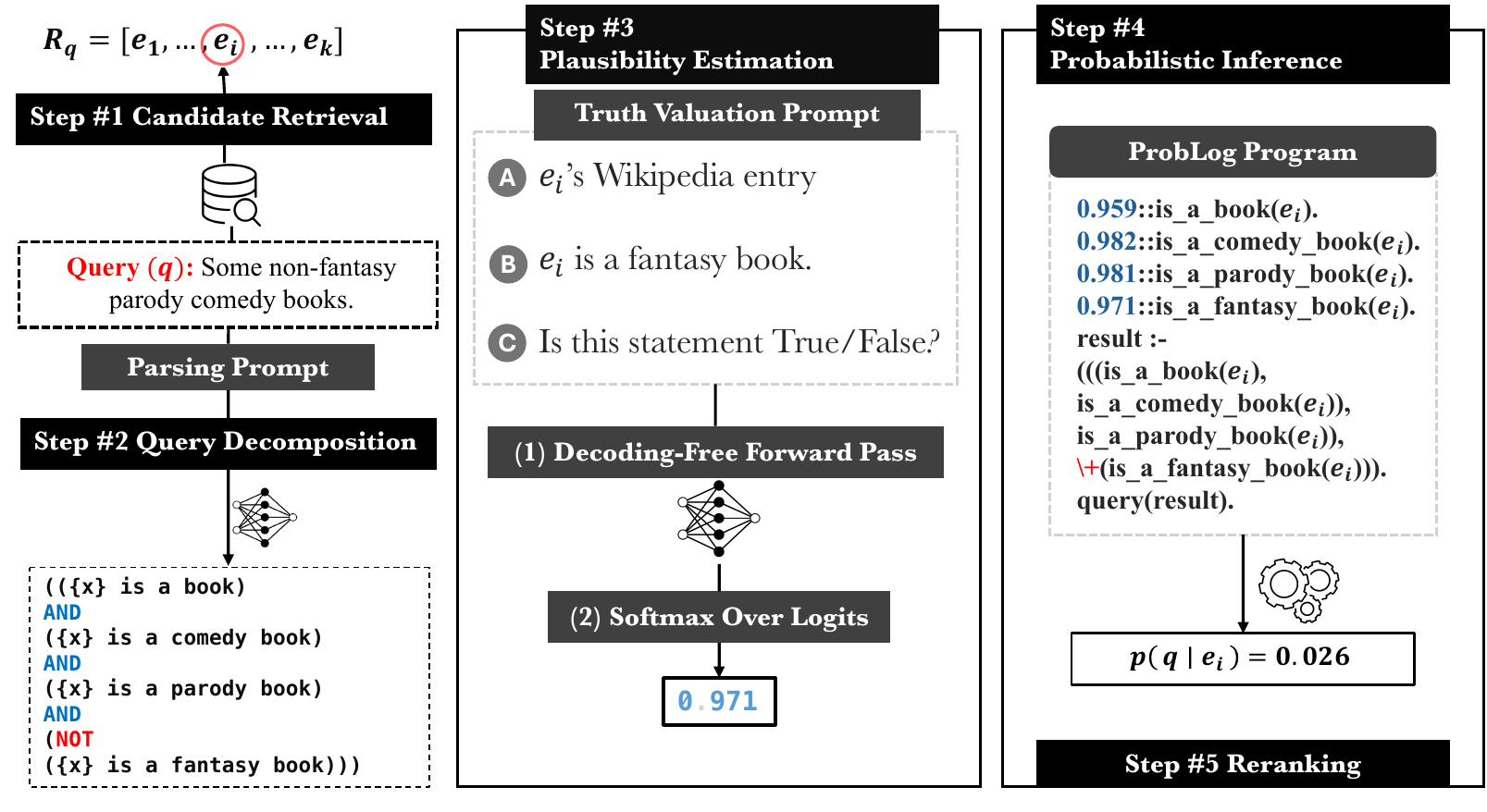} 
  \caption{Overview of \textbf{OrLog}, a neuro-symbolic framework for resolving complex entity-seeking queries. For a given query five steps are followd.
\textit{Step~1}: a retriever narrows the search space to candidate entities $(R_q)$. 
\textit{Step~2}: a semantic parser decomposes the query into atomic predicates and a logical form. 
\textit{Step~3}: predicate priors are elicited from an LLM via a truth-valuation prompt, using a single decoding-free forward pass. 
\textit{Step~4}: these priors instantiate a probabilistic program in ProbLog, which computes posterior query satisfaction for each entity. 
\textit{Step 5}: Candidates are reranked by their inferred probabilities.}
  \label{fig:framework_overview}

\end{figure*}

\section{Introduction}

\vspace*{-2mm}
Many real-world information needs involve multiple interacting constraints~\cite{10.1145/24634.24635, DBLP:conf/acl/MalaviyaSCLT23}. Consider, e.g., the query \textit{``Museums in Paris or Amsterdam but not art museums''}, which requires not only identifying semantically related entities but also enforcing how conjunction, disjunction, and negation restrict the answer space~\cite{weller-etal-2024-nevir, DBLP:conf/aaai/ZhangZ0PRRCR25}. In formal terms, a valid entity $e$ must satisfy $\mathit{museum}(e)\;\wedge\;(\mathit{paris}(e)\;\vee\;\mathit{amsterdam}(e))\;\wedge\;\neg\mathit{art}(e)$. Resolving such queries involves evaluating whether candidate entities satisfy a logical constraint, rather than merely ranking by semantic similarity.
Current retrieval systems are ill-equipped 
for addresssing such queries~\cite{van_den_Elsen_2025, DBLP:conf/acl/MalaviyaSCLT23, DBLP:conf/nlpir/MaiG024}. Sparse and dense models approximate relevance by lexical overlap or embedding similarity. Such relevance scores do not compose in ways that respect logical operators: the outcome for logical constraints over a query’s atomic predicates cannot necessarily be derived from token- or embedding-level similarities between a document and predicates~\cite{weller2025theoreticallimitationsembeddingbasedretrieval,petcu2025comprehensive,10.1145/24634.24635}.

 End-to-end reasoning with LLMs that attempts to enforce constraints directly in token space~\cite{DBLP:conf/nips/Wei0SBIXCLZ22} can also be unreliable and computationally expensive~\cite{arcuschin2025chainofthoughtreasoningwildfaithful,yue2025does}, despite being capable of producing plausible answers. Neuro-symbolic approaches could, in principle, enforce constraints faithfully and provide a viable path for modeling reasoning in complex retrieval settings. However, existing systems assume unambiguous queries and complete evidence at inference time, conditions that may fit logical puzzles and mathematical problems, but are rarely satisfied in information retrieval~\cite{Olausson_2023, yang2023leandojo,DBLP:conf/emnlp/PanAWW23}.

\ourparagraph{Approach.}
We introduce \textbf{OrLog}, a neuro-symbolic retrieval framework that decouples \emph{predicate-level plausibility estimation} from \emph{logical reasoning}. OrLog uses a standard retriever to narrow the search space, then invokes an LLM to decompose the natural language query into atomic predicates and a logical form. 
Rather than performing end-to-end reasoning with an LLM, in our framework, the LLM acts as an \emph{uncertainty oracle}: given an entity and a predicate, the model returns a plausibility score in a single forward pass without free-form generation. This score is a scalar between 0 and 1 obtained by transforming the model’s logits for designated truth labels; it quantifies the model’s degree of support that the predicate holds for the entity in the context of the query. 
These plausibility scores are then used as predicate-level priors in \textsc{ProbLog}~\cite{DBLP:conf/ijcai/RaedtKT07}, to compute the posterior probability of an entity satisfying the full query. By shifting compositional reasoning to a probabilistic reasoning engine, OrLog enforces constraints efficiently, avoiding the verbosity and unreliability of monolithic LLM reasoning. Our experiments address three research questions:
\begin{enumerate}[label=(RQ\arabic*),leftmargin=*,nosep]
\item How does \textit{OrLog} compare to \textit{LLM-as-reasoner}  on resolving complex set-compositional queries?
\item How does the choice of the underlying LLM affect OrLog's performance and that of the LLM?
\item How does query structure (e.g., conjunction, disjunction, and negation) affect the relative performance of OrLog and monolithic LLMs?
\end{enumerate}

\ourparagraph{Contributions.} The four main contributions of this work are:
\begin{enumerate}[leftmargin=*,nosep]
    \item A neuro-symbolic framework, OrLog, that combines retrieval, LLM-based predicate plausibility estimation, and probabilistic reasoning for constraint-aware re-ranking for complex queries. 
    \item An efficient predicate-level plausibility estimation
 method that elicits plausibility scores from LLM logits in a decoding-free forward-pass.
    \item Empirical evidence that OrLog achieves significant performance gains over 
    LLM-as-reasoner while reducing token usage by $\sim$90\%.
    \item A structural analysis of how logical operators in queries affect the comparative performance of OrLog versus monolithic LLM reasoning. 
\end{enumerate}

\vspace*{-2mm}
\section{Related Work}

\vspace*{-2mm}
\subsection{Logic-based information retrieval}
\vspace*{-1mm}
The `logical approach to information retrieval (IR)' is an early line of research to ground retrieval models in formal axioms and logical theories~\cite{10.1145/24634.24635, DBLP:conf/sigir/BruzaH94}, taking the stand that logic would provide a necessary conceptual foundation for modeling retrieval~\cite{DBLP:journals/jasis/RijsbergenL96}. Documents and queries would be represented as logical sentences, relevance defined by logical implication~\cite{DBLP:journals/sigir/HuibersLR96}, and uncertainty through logical imaging~\cite{Crestani1998}. While theoretically rigorous, these frameworks faced practical obstacles in their application across large collections. Also, relative performance gains (if any) over previous approaches never outweighed their increased system and modeling complexity. As a result, the pursuit of logically grounded retrieval systems has remained unresolved~\cite{DBLP:conf/trec/CrestaniRSR95, DBLP:journals/csur/AbdulahhadBCP19}.

\vspace*{-2mm}
\subsection{Reasoning: Monolithic and LLM-modulo solutions}
\vspace*{-1mm}
Contemporary approaches to complex, logic-intensive natural language tasks increasingly rely on LLMs as general-purpose monolithic reasoners.
These models are incentivized to produce verbose sequences of intermediate steps, so-called chain-of-thought (CoT), to approximate structured reasoning~\cite{DBLP:conf/nips/Wei0SBIXCLZ22}. While this strategy often improves final accuracy, it provides no guarantee of logical coherence: the generated chain may obscure inconsistencies in intermediate steps or rationalize an incorrect answer~\cite{lanham2023measuring, arcuschin2025chainofthoughtreasoningwildfaithful, yee2024faithful}. Reinforcement learning techniques that promote longer traces amplify the risk of confabulated content~\cite{10.1145/3703155, yue2025does}, as seen in recent OpenAI models compared to earlier ones~\cite{o3o4mini_system_card_2025, kalai2025languagemodelshallucinate}. Beyond the concerns of faithful reasoning, verbose generation incurs non-trivial computational cost, particularly in settings requiring extended or branching reasoning paths~\cite{DBLP:journals/corr/abs-2504-10903, DBLP:journals/corr/abs-2506-02153}. An alternative line of work externalizes reasoning: LLMs act as stochastic components that excel in semantic understanding and parsing, while symbolic backends enforce structured reasoning~\cite{deepmind2024imo}. Within this LLM-modulo paradigm~\cite{kambhampati2024position, Hoveyda_2025}, systems diverge in the choice of backend. LINC~\cite{Olausson_2023} employs a first-order prover, achieving high logical fidelity for logical puzzles but requiring exact translations and complete information at inference time. Some cognitive science-inspired approaches to modeling reasoning extend this principle further, casting LLMs as front-ends that synthesize bespoke probabilistic programs, with reasoning delegated to probabilistic program interpreters to support reasoning in novel situations~\cite{DBLP:journals/corr/abs-2507-12547, wong2023wordmodelsworldmodels}. Closer to our setting, Nafar et al.~\cite{DBLP:conf/aaai/NafarVK25} use probabilistic logic programming as a principled substrate for uncertainty, outperforming LLMs in probabilistic reasoning but assuming predicate probabilities are given.

\vspace*{-2mm}
\subsection{LLM uncertainty estimation}
\vspace*{-1mm}
Since OrLog treats the LLM as a probabilistic oracle, its effectiveness hinges on how reliably we can estimate predicate-level uncertainty from the model. UE methods for LLMs can be divided into white-box and black-box approaches~\cite{DBLP:journals/corr/abs-2412-05563}. White-box methods exploit internal logits to quantify predictive uncertainty~\cite{Kadavath22PE,Bakman24mars,Duan24SAR,Kuhn23SE}. Despite access to logits, these methods typically rely on multiple decoding passes or sampled generations to approximate uncertainty over full responses, leading to high computational cost. Black-box methods treat the model as an opaque generator, inferring uncertainty from similarity or dispersion across sampled outputs~\cite{Ng01Advances,Lin24ECC}. Both paradigms assume free-form generation tasks and become unstable with suboptimal external evidence, as in retrieval-augmented generation (RAG)~\cite{soudani-etal-2025-uncertainty}.
OrLog departs from both lines as predicate-level uncertainty is elicited in a decoding-free setting. For each predicate, a single forward pass yields the logits for tokens \texttt{True}/\texttt{False}, which are normalized into a final score, indicating the truth probability of that predicate. This approach avoids repeated sampling, reduces token cost, and yields uncertainty estimates tailored for probabilistic logical retrieval.

\vspace*{-2mm}
\section{The OrLog Reasoning Framework}

\vspace*{-2mm}
\subsection{Problem definition}
\vspace*{-1mm}
Let \(\mathcal{E}\) be the corpus of entities, where each entity is characterized by a textual description and an identifier.
The textual representation of an entity \(e \in \mathcal{E}\) may be a document (e.g., a Wikipedia article) or a set of (semi-)struc\-tured facts (e.g., a Wikidata item). We focus on Wikipedia documents.
The user's natural-language query \(q\) implicitly specifies logical constraints over \(\mathcal{E}\). E.g., the query \textit{``Some non-fantasy parody
comedy books''} contains atomic predicates $A$, $B$, $C$, and $D$ combined through formal connectives as $A \cap B \cap C \cap \neg D$, as illustrated in Fig.~\ref{fig:framework_overview}. 
Our goal is to return a ranked list of entities $E = [e^*_1, e^*_2, \ldots, e^*_n]$ that contains relevant entities to the query \(q\), satisfying the logical constraints expressed in the query.

\vspace*{-2mm}
\subsection{Framework overview}
\vspace*{-1mm}
We propose a neuro-symbolic framework that combines LLMs and formal probabilistic logic to improve trustworthiness and retrieval effectiveness.  
Fig.~\ref{fig:framework_overview} sketches our \textit{neuro-symbolic retrieval} framework, \textbf{OrLog}, in five steps: 
(i) \textit{Candidate retrieval} retrieves the top-$k$ entities via a sparse or dense model. 
(ii) \textit{Query decomposition} extracts atomic predicates from the natural‐language query and constructs its Boolean logical representation. 
(iii) \textit{Predicate Plausibility estimation} estimates plausibility scores for atomic predicates. 
(iv) \textit{Probabilistic reasoning} computes the posterior score of an entity satisfying the query over the weighted predicates.
(v) \textit{Reranking} sorts the candidates by their posterior probabilities in descending order. 
Below, we detail each component.

\vspace*{-2mm}
\subsection{Candidate  retrieval}
\vspace*{-1mm}
Reasoning over all entities in the entity corpus \(\mathcal{E}\) for each incoming query $Q$ is intractable. We therefore narrow down the search space to the top-$k$ ranked results using a base retrieval model. This cost-effective filtering ensures that the subsequent reasoning procedure is applied only to a promising list of candidate entities \(\mathcal{R}_q\), likely to contain entities relevant to the query.

\vspace*{-2mm}
\subsection{Query decomposition}
\vspace*{-1mm}
We next derive a \textit{formal representation} of natural language query $q$ to enable symbolic reasoning over its candidate entities, by decomposing $q$ into a finite set of atomic predicates 
$\mathcal{P} = \{P_1, P_2, \ldots, P_n\}$. Each predicate is a function of entity $e$ that represents a factual statement, taking the value of \textsc{True} or \textsc{False}. For example, predicate $P_1(e) = {}$``$\{e\}$ is a comedy book'' is applied to every candidate entity (illustrated in Fig.~\ref{fig:framework_overview}). These predicates are combined using Boolean connectives (conjunction \textsc{and}, disjunction \textsc{or}, and negation \textsc{not}) to represent the constraints of the query. Fig.~\ref{fig:framework_overview} illustrates how the formal representation of query \textit{``Some non-fantasy parody comedy books''} consists of four predicates combined with logical connectives \textsc{and} and \textsc{not}.

To perform the transformation from a natural language query to a formal proposition, an LLM is prompted with a one-shot example. The LLM produces: (i) a set of atomic predicates \( \mathcal{P} \)  extracted from query \( q \), and (ii) a logical proposition of the query  \( q' \), such that \( q' \) approximates the logical semantics of \( q \). The fidelity of this translation directly affects downstream reasoning in OrLog. In the absence of ground-truth translations, it is difficult to quantify translation accuracy and its effect on OrLog's performance, but manual inspection of a sample of queries suggests that the translated representations are of satisfactory quality.%
\footnote{%
We do not experiment with the choice of LLM for this step, using \emph{Llama-3.3-70B-instruct} for all transformations in our experiments.}

\vspace*{-2mm}
\subsection{Predicate plausibility estimation}
\vspace*{-1mm}
In probabilistic logic, every predicate (or atom) is annotated with a prior probability to indicate the likelihood of the predicate being true. Here, we introduce our method to estimate these scores. We utilize the logits produced by an LLM for the tokens \textit{True} and \textit{False} in response to a truth-valuation prompt. The plausibility score, then, is a scalar value between 0 and 1, indicating model’s degree of support that the predicate holds true for the entity. While we will treat these scores as prior probabilities in the downstream probabilistic program, we emphasize that they originate as plausibility estimates of predicate truth, i.e., calibrated degrees of support in $[0,1]$ rather than probabilities in the strict axiomatic sense~\cite{Polya1968-PLYMAP, Jaynes2002-JAYPTT}.

\ourparagraph{Truth valuation prompt.} Given an entity $e$, we construct a truth-valuation prompt $T= \langle c, t_e, P(e), \sigma\rangle$ consisting of four elements: 
(i) context $c$, which is a description of the entity (e.g., from Wikipedia); 
(ii) the entity title $t_e$; 
(iii) the atomic predicate $P(e)$, indicating a factual statement about entity $e$; and 
(iv) a truth inquiry suffix $\sigma$ (e.g., “Is this predicate True or False?”) to facilitate eliciting LLM’s probability output. The predicate and entity title are mandatory parts of the prompt, the rest is optional. These elements are concatenated and fed into an LLM in a single forward pass.

\ourparagraph{Computing plausibility score.}
LLM’s estimated plausibility that the atomic predicate \(P(e)\) holds is calculated as the softmax score assigned to the token \textit{True}. 
Let \(z_{True},z_{False}\) be the LLM final layer's logits for the tokens \textit{True} and \textit{False}, respectively, given the truth valuation prompt $T$.
Following prior work that derives relevance estimates from LLMs' logits over target tokens \cite{zhuang-etal-2024-beyond, nogueira-etal-2020-document}, here we estimate the truth plausibility of each atomic predicate from the LLM as:
\[
\pi(P(e)) \;=\; \frac{\exp(z_{\mathrm{True}})}{\exp(z_{\mathrm{True}}) + \exp(z_{\mathrm{False}})}.
\]

\vspace*{-2mm}
\subsection{Probabilistic reasoning}
\vspace*{-1mm}
For each $e$ in $\mathcal{R}_q$, we assemble a ProbLog program that includes
the weighted predicates \(\{ \langle \pi(P(e)), P(e) \rangle \mid P\in\mathcal{P}\}\)
and the logical proposition of the query $q'$.
ProbLog reasoning then computes the posterior probability of \(e\) satisfying the query $e$, denoted as $\rho(q, e)$. 
ProbLog interprets the program as a distribution over truth assignments
to the predicates of \(e\), and returns the posterior probability
\(\rho(q,e)\) by summing the weights of all assignments in which \(q'\) holds.
For conjunctive queries, this reduces to a product of predicate probabilities,
while disjunctive and mixed queries require marginalization over multiple
satisfying worlds. As shown in step 4 in Fig.~\ref{fig:framework_overview}, this yields the
final score used for reranking.

\vspace*{-2mm}
\subsection{Reranking}
\vspace*{-1mm}
Finally, candidates are ranked in descending order of their posterior probability \(\rho(q,e)\), yielding the final list \([e_{(1)}, \dots, e_{(k)}]\) (final step in Fig.~\ref{fig:framework_overview}).
In cases where two entities receive identical posteriors, ties are resolved by preserving their relative order from the originally retrieved ranking.

\vspace*{-2mm}
\section{Experiments}

\vspace*{-2mm}
\subsection{Dataset}
\vspace*{-1mm}
Evaluating whether retrieval systems can resolve complex queries with logical constraints requires test queries that go beyond topical or semantic relevance and involve interacting logical constraints expressed in natural language. The QUEST dataset~\cite{DBLP:conf/acl/MalaviyaSCLT23} is well-aligned with this need: it consists of queries mapped to sets of Wikipedia entities defined through set-theoretic operations, intersection, union, and difference, and their combinations. Queries are derived from Wikipedia category compositions (e.g., \textit{science-fiction films shot in England}) and span seven templates, covering multiple constraint types and logical complexities. We run experiments on the 1,727 query test split released by the authors.

\vspace*{-2mm}
\subsection{Retrievers}
\vspace*{-1mm}
For candidate retrieval, the first step of OrLog, we use two models to retrieve the top-$k$ entities per query: \textit{BM25}~\cite{DBLP:journals/ftir/RobertsonZ09} as a sparse lexical model, and \textit{E5-base-v2}~\cite{Wang2022TextEB} as a dense model optimized for semantic similarity that is shown to perform reasonably well on QUEST~\cite{shen-etal-2025-logicol}. We set $k = 20$ in all experiments to ensure a tractable pool for downstream reasoning. Out of the 1,727 test queries in QUEST, BM25 retrieves at least one gold entity for 644 queries, while E5 does so for 981 queries. Our reported results are based on the complete test set in QUEST.

\vspace*{-2mm}
\subsection{LLMs}
\vspace*{-1mm}
In our work, LLMs are used in three distinct roles:
\begin{enumerate}[leftmargin=*]
    \item \textbf{Reasoning} Serving as \textit{LLM-as-reasoner} baseline, the LLM receives the query and entity information and is prompted to decide whether each entity satisfies the query, and to output the final result in True/False.
    \item \textbf{Translation} Parsing the natural-language query into atomic predicates and logical forms, which will later be used in OrLog.
    \item \textbf{Plausibility estimation} eliciting predicate-level plausibility scores later assigned as priors to weighted predicates in ProbLog.
\end{enumerate}
For translation, we fix the model across all experiments to \textit{Llama-3.3-70B-Instruct} to isolate the variability in parsing quality. 
Both the LLM-as-reasoner and OrLog frameworks are evaluated across three backbone LLM families: \textit{Mis\-tral-v1} (7B-Instruct,  8×7B-Ins\-truct), \textit{Qwen-2.5} (7B-Instruct, 72B-Instruct), and \textit{Llama-3} (8B-Instruct, 70B-Instruct). These models vary in scale, architecture, and training approach, allowing us to analyze the effect of different model families and sizes on both predicate plausibility estimation and monolithic LLM reasoning. 
Due to limited local compute resources, we offload LLM reasoning experiments to the OpenRouter API rather than hosting large LLMs on-premises. Since OpenRouter does not provide access to model logits, we perform all OrLog-related experiments on our H100 or A100 GPU instances, wherever relevant. Larger backbones are 4-bit quantized to fit within a single GPU.
\footnote{Our repository containing the implementation, prompt templates, and experiments is available at \href{https://github.com/informagi/RSN}{github.com/informagi/RSN}.}

\vspace*{-2mm}
\subsection{Knowledge availability setups}
\vspace*{-1mm}
To examine the impact of information access during reasoning process in each method, we evaluate them under two access levels:
\begin{enumerate*}[label=(\roman*), itemjoin={; }, itemjoin*={; and  }]
  \item \textbf{Parametric}: only the entity name and the query are provided; the LLM relies exclusively on its internal (parametric) knowledge to estimate predicate truth plausibility or assess query satisfaction
  \item \textbf{Parametric+}: in addition to the query and entity name, we provide a Wikipedia description of the candidate entity (explicit external evidence). This text may or may not contain a direct answer, but is intended to elicit more relevant parametric knowledge in OrLog’s setup or to enhance reasoning in the LLM-as-reasoner setup
\end{enumerate*}.
This comparison allows us to assess how the absence of contextual information affects performance across frameworks.
\vspace*{-2mm}
\subsection{Baselines}

\vspace*{-1mm}
We evaluate three baseline configurations. The first is the \textbf{Retriever-Only} setup, where we directly evaluate the rankings produced by the base retrievers without applying any further reasoning. The second is the \textbf{LLM-as-reasoner} setup, where LLMs are prompted in an end-to-end fashion with both the query and candidate entity information (\ensuremath{\pm} evidential knowledge) to decide whether each entity satisfies the query. The third baseline, \textbf{Logic-Augmented LLM}, is based on the LLM-as-reasoner setup, with the prompt being augmented by the logical decompositions of the query to help the LLM reasoning process. These decompositions are the ones used in OrLog, and the motivation for adding this baseline is to have comparative setups between OrLog and LLM reasoning.

\vspace*{-2mm}
\subsection{Evaluation metrics}
\vspace*{-1mm}
We evaluate retrieval performance using Precision@K (P@K), Recall@K (R@K), F1@K, and NDCG@K at $K\in\{1,10\}$, as well as Mean Reciprocal Rank (MRR).
All metrics are averaged over the test set and, where relevant, reported per query type.
To assess statistical significance, we perform paired two-tailed sign tests and report the \textit{p}-values.
For cost evaluation, we compare the systems in terms of the number of tokens generated by LLMs. For the \textit{LLM-as-reasoner} setup, we measure the total number of tokens generated by the language model across all query–entity pairs and report the average per pair. For \textit{OrLog}, the token count per query–entity pair comprises (1) the number of atomic predicates in the parsed query, since each predicate triggers one forward pass to the LLM, and (2) the one-off token cost to produce the parsing. Final average is reported over all query–entity pairs.

\vspace*{-2mm}
\section{Experimental Results}

\vspace*{-1mm}
We answer our questions from the introduction: 
(i) Does isolating predicate-level uncertainty to an LLM and delegating reasoning to ProbLog yield higher retrieval accuracy than monolithic LLM reasoning? 
(ii) How does the choice of backbone LLM affect reasoning quality in both OrLog and LLM-as-reasoner frameworks? 
And (iii) which classes of logical constraints most clearly reveal the performance gap between OrLog and LLM-as-reasoner approaches? We examine each question in sequence below.

\vspace*{-2mm}
\subsection{OrLog vs.\ LLM for complex query resolution}

\vspace*{-1mm}
Table~\ref{tab:performance} presents the evaluation results for OrLog and the LLM-as-reasoner baseline across two base retrievers (BM25 and E5) and two knowledge access settings (\textit{Parametric} and \textit{Parametric+}).

\newcommand{\KaiFu}[1]{\multicolumn{1}{c}{#1}}
\newcommand{\sigbullet}{\textsuperscript{$\bigcirc$}}     %
\newcommand{\sigbox}{\textsuperscript{\ensuremath{\Box}}}
\newcommand{\sigtriangle}{\textsuperscript{$\triangle$}} %
\newcommand{\regParam}{\textsuperscript{$\lozenge$}}        %
\newcommand{\regParamPlus}{\textsuperscript{$\blacklozenge$}} %

\begin{table*}[t]
  \small
  \centering
  \shrink
  \caption{Results for BM25 and E5 base retrievers (shaded) and reranked variants using \textit{Llama-3.3-70B-Instruct} across three reasoning strategies: \textbf{LLM}, \textbf{Logic-augmented LLM} (LLM*), and \textbf{OrLog}. 
  Significance of OrLog (better or worse) is tested against 
the comparable LLM baseline (Param or Param+);\sigbullet $p{<}0.05$, $\Box$ $p{<}0.01$, $\triangle$ $p{<}0.001$.
  Bold indicates the best score per base retriever and metric. }

  \setlength{\tabcolsep}{2pt}  
  \renewcommand{\arraystretch}{1.1} 
  \resizebox{\textwidth}{!}{
  \begin{tabular}{l ll ll ll ll l}
    \toprule
    \textbf{System}
      & \multicolumn{2}{c}{\textbf{P@K}}
      & \multicolumn{2}{c}{\textbf{R@K}}
      & \multicolumn{2}{c}{\textbf{F1@K}}
      & \multicolumn{2}{c}{\textbf{NDCG@K}}
      & \textbf{MRR} \\ 

    \cmidrule(lr){2-3} \cmidrule(lr){4-5}
    \cmidrule(lr){6-7} \cmidrule(lr){8-9}
    & \KaiFu{1} & \KaiFu{10}
    & \KaiFu{1} & \KaiFu{10}
    & \KaiFu{1} & \KaiFu{10}
    & \KaiFu{1} & \KaiFu{10}
    & \\
    \midrule

    \rowcolor{gray!15} \textbf{BM25} 
      & 0.116 & 0.067 & 0.014 & 0.073 & 0.022 & 0.062 & 0.116 & 0.089 & 0.175  \\
  + LLM*(Param)       
      & 0.158 & 0.083 & 0.019 & 0.088 & 0.031 & 0.077 & 0.158 & 0.114  & 0.219 \\
    + LLM(Param)       
      & 0.155 & 0.079 & 0.020  & 0.085 & 0.032 & 0.074 & 0.155 & 0.110  & 0.212 \\

    + OrLog(Param)      
      & 0.138\sigbox & 0.075 & 0.015  & 0.080\sigbox & 0.026 & 0.070\sigbox & 0.138 & 0.101\sigbullet & 0.199\\
      + LLM*(Param+)      
      & 0.144 & 0.076 & 0.019 & 0.083 & 0.030 & 0.071 & 0.144 & 0.106  & 0.203 \\
      + LLM(Param+)    
      & 0.185 & 0.086 & 0.023 & 0.093 & 0.038 & 0.080 & 0.185 & 0.125 & 0.244 \\
    + \textbf{OrLog(Param+)}       
      & \textbf{0.221}\sigtriangle & \textbf{0.089}\sigbullet 
      & \textbf{0.026}\sigtriangle  & \textbf{0.095}\sigbullet 
      & \textbf{0.044}\sigtriangle & \textbf{0.082}\sigbullet 
      & \textbf{0.221}\sigtriangle & \textbf{0.134}\sigtriangle 
      & \textbf{0.267}\sigbox \\

    \rowcolor{gray!15} \textbf{E5} 
      & 0.182 & 0.105 & 0.021 & 0.116 & 0.035 & 0.098 & 0.182 & 0.142 & 0.269 \\
        + LLM*(Param)       
      & 0.204 & 0.119 & 0.024 & 0.132 & 0.041 & 0.111 & 0.204 & 0.163  & 0.299 \\
    + LLM(Param)       
      & 0.192 & 0.115 
      & 0.023 & 0.126
      & 0.038 & 0.107
      & 0.192 & 0.156
      & 0.287 \\

    + OrLog(Param)     
      & 0.161\sigbox & 0.104\sigtriangle & 0.019\sigbox  & 0.114\sigtriangle  & 0.031\sigbox & 0.097\sigtriangle  & 0.161\sigbox & 0.136\sigtriangle  & 0.260\sigbox \\
         + LLM*(Param+)       
      & 0.241 & 0.127 & 0.031 & 0.138 & 0.050 & 0.118 & 0.241 & 0.180  & 0.333 \\
      + LLM(Param+)    
      & 0.263 & \textbf{0.133} & 0.034 & \textbf{0.145} & 0.056 & \textbf{0.123} & 0.263 & 0.193 & 0.355 \\

    + \textbf{OrLog(Param+)}       
      & \textbf{0.308}\sigtriangle & 0.131 
      & \textbf{0.039}\sigtriangle & 0.142 
      & \textbf{0.063}\sigtriangle & 0.121 
      & \textbf{0.308}\sigtriangle & \textbf{0.197}\sigbullet 
      & \textbf{0.384}\sigbox  \\
    \bottomrule
  \end{tabular}
  }

  \label{tab:performance}
\end{table*}

\ourparagraph{Effect of access to external knowledge.}
We notice at the first glance that OrLog\textit{(Param+)} significantly outperforms all baselines with respect to all metrics in all setups, except P, R, and F1 at rank 10 with the E5 retriever.  In the \textit{Parametric} setting, where only the entity name is provided, OrLog significantly underperforms the LLM-as-reasoner, with the largest deficit under E5. Lacking contextual grounding, the LLM’s predicate plausibility estimates are poorly calibrated; \textsc{ProbLog} receives miscalibrated priors and performance degrades. The pattern also demonstrates that OrLog gains more from external evidence than LLM-as-reasoner, suggesting more effective integration of factual grounding into the reasoning process. Interestingly, the \textit{Logic-Augmented LLM} experiments shows that augmenting the prompt with atomic decompositions and a logical form does not consistently improve performance. While such structure is intended to guide more faithful inference, in the \textit{Param+} setting it leads to degradation relative to the plain LLM. A likely cause is that the additional instructions and representations introduce redundancy and cognitive load for the model, which struggles to integrate them effectively. In contrast, decomposition is one of the primary sources of OrLog’s performance gains.

\ourparagraph{Effect of the base retriever.}
The gains with BM25 are more uniform than with E5.
While in the \textit{parametric+} setup, OrLog substantially outperforms the LLM baseline on $\mathrm{NDCG}$, its gains are less consistent in other metrics. A likely cause is our fallback mechanism: when the LLM-as-reasoner fails to emit a valid binary label, we revert to the base-retriever score. Because E5 already yields strong initial rankings, this disproportionately benefits the baseline. Still, OrLog shows sharper top-rank precision; consistently higher P@1 and NDCG@1. Overall, improvements correlate with retriever weakness: OrLog delivers larger gains with BM25 and only modest lift on an already-strong E5.

\ourparagraph{Cost comparison.}
Beyond ranking efficacy, practical deployment demands reasoning efficiency as well. Table \ref{tab:avg_tokens} reports the mean token count per query–entity invocation; end-to-end LLM reasoning uses \(35\!-\!55\) tokens on average (higher under \textit{Parametric+} due to longer prompts and outputs), whereas OrLog remains below \(6\) tokens. This reduction stems from (i) decoding-free predicate plausibility estimation (a single forward pass per predicate) and (ii) one-off query parsing with composition delegated to \textsc{ProbLog}, whose runtime and computational cost are negligible relative to an LLM call.
\begin{table}[t]
  \centering
  \shrink
    \caption{Average generated token count per query-entity for each method.OrLog reduces token generation by $\sim$90\% compared to LLM-as-reasoner approaches.}
  \small
  \setlength\tabcolsep{3pt}    
  \begin{tabular}{llc}
    \toprule
    Retriever & Method & \makecell{Avg. generated tokens} \\
    \midrule
    \multirow{3}{*}{BM25}
      & LLM (Param)  & 50.2 \\ 
      & LLM (Param+) & 55.0 \\ 
      & \textbf{OrLoG}        & \textbf{5.80} \\ 
    \midrule
    \multirow{3}{*}{E5}
      & LLM (Param)  & 35.3 \\ 
      & LLM (Param+) & 46.6 \\ 
      & \textbf{OrLoG}        & \textbf{5.79} \\ 
    \bottomrule
  \end{tabular}
  \shrink
  \label{tab:avg_tokens}
\end{table}

\vspace*{-2mm}
\subsection{Effective predicate plausibility estimation may offset model size}
\vspace*{-1mm}
Table~\ref{tab:llm} depicts the performance variation of OrLog and the LLM-as-reasoner pipeline with various backbone models. In the parametric knowledge setting, LLM-as-reasoner consistently outperforms OrLog across all families of models (Mistral, Qwen, and Llama). This, again, confirms that, absent any grounding information, predicate plausibility estimates from the LLM are too noisy to support robust probabilistic reasoning, causing OrLog’s rankings to lag.
Following the same trend in Table~\ref{tab:performance}, introducing Wikipedia descriptions (+Parametric) boosts both frameworks, but OrLog benefits more effectively from the provided context. In this table, the Qwen‐2.5-7B results introduce an important nuance: despite its modest scale, the small Qwen backbone supplies predicate estimates of sufficient fidelity to enable OrLog to outperform even the 72B variant. Concretely, OrLog (Parametric+) with Qwen-7B achieves P@1 = 0.303 and MRR = 0.381, slightly surpassing both its large-model OrLog counterpart (0.300/0.374) and the monolithic LLM 72 B (0.294/0.380). This indicates that various factors, such as training and post-training quality, can compensate for parameter count when generating calibrated plausibility estimates. 

\vspace*{-2mm}
\subsection{Disjunction amplifies OrLog's gains over LLM-based reasoning}
\begin{figure*}[t]
\shrink
  \centering
  \includegraphics[trim=0 20 0 0, clip, width=1\textwidth]{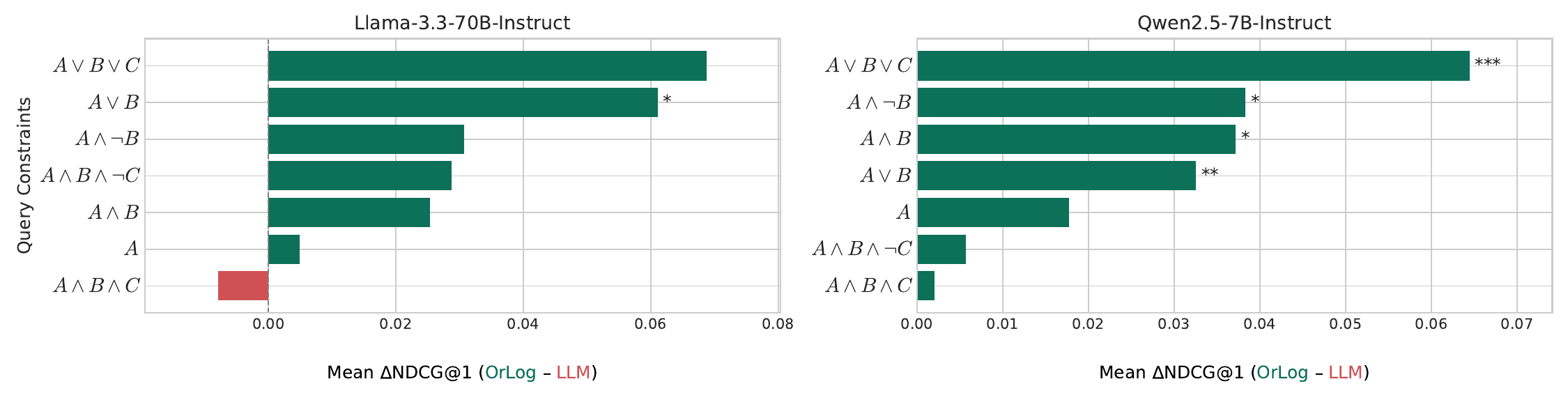} 
  \shrink
  \caption{Mean $\Delta$P@1 between \textit{OrLog} and \textit{LLM-as-reasoner} baseline across query‐template structures in QUEST, in the \textit{informed} knowledge access setting. Each bar shows the average per-query gain or loss for a specific template, with green indicating OrLog outperforming the LLM-as-reasoner baseline, and red indicating the reverse.  Left and right panels correspond to different underlying LLMs: \textit{Llama-3.3-70B-Instruct} and \textit{Qwen-2.5-7B-Instruct}. Larger deviations highlight structural patterns where symbolic probabilistic reasoning offers a clear advantage relative to reasoning purely with LLMs.}
  \label{fig:gain_across_templates}
\end{figure*}

\vspace*{-1mm}
Figure~\ref{fig:gain_across_templates} depicts mean $\Delta$P@1 between OrLog and LLM-reasoner, stratified by query structural template taken from QUEST, evaluated on Llama-3.3-70B and Qwen-2.5-7B. Three clear patterns emerge. First, OrLog’s benefit is negligible for purely \textit{conjunctive} queries, as for $A\land B$ and $A\land B\land C$, gains are negligible (typically $\approx0.02$ or lower), indicating that well-prompted LLMs compose simple conjunctive constraints nearly as well as a symbolic backend if not better.

\tikzset{marker style/.style={line width=0.8pt, fill opacity=0.95}}

\newlength{\markerwidth}
\setlength{\markerwidth}{1.5em}
\setlength{\lightrulewidth}{0.03em}

\newsavebox\SMALLBASEBOX
\savebox\SMALLBASEBOX{%
  \tikz[baseline=-0.5ex]{%
    \draw[marker style,draw=PS2,fill=white] (0,0) circle (0.6ex);
  }%
}
\newcommand{\SmallBaseMarker}{\usebox\SMALLBASEBOX}

\newsavebox\SMALLBOX
\savebox\SMALLBOX{%
  \tikz[baseline=-0.5ex]{%
    \draw[marker style,draw=PS2,fill=PS2] (0,0) circle (0.6ex);
  }%
}
\newcommand{\SmallMarker}{\usebox\SMALLBOX}

\newsavebox\LARGEMARKERBOX
\savebox\LARGEMARKERBOX{%
  \tikz[baseline=-0.5ex]{%
    \draw[marker style,draw=PS3,fill=PS3] (0,0) circle (0.8ex);
  }%
}
\newcommand{\LargeMarker}{\usebox\LARGEMARKERBOX}

\newsavebox\LARGEMARKERBASEBOX
\savebox\LARGEMARKERBASEBOX{%
  \tikz[baseline=-0.5ex]{%
    \draw[marker style,draw=PS3,fill=white] (0,0) circle (0.8ex);
  }%
}
\newcommand{\LargeBaseMarker}{\usebox\LARGEMARKERBASEBOX}

\newcommand{\sizehdr}[1]{\multicolumn{3}{c}{#1}}
\newcommand{\metahdr}{\text{@1} & \text{@10} & \text{MRR}}

\newcommand{\sigspace}{\mkern1mu}
\sisetup{add-integer-zero = false}

\begin{table*}[t]
  \centering
  \renewcommand{\arraystretch}{1.10}
  \shrink
    \caption{Effect of various backbone LLMs on the performance of OrLog vs.\ LLM-as-reasoner with E5 as the candidate retriever. Each panel reports results for two size variants of Mistral-v1, Qwen-2.5, and Llama-3. 
Significance of OrLog (better or worse) is tested against 
the comparable LLM baseline (Param or Param+);\sigbullet $p{<}0.05$, $\Box$ $p{<}0.01$, $\triangle$ $p{<}0.001$.
Boldface denotes the best score.}%
\miniskip
\begin{subtable}[t]{0.485\textwidth}
  \centering
  \caption{Mistral-v1}
  \miniskip
  \resizebox{\linewidth}{!}{%
    \begin{tabular}{@{} l S S S S S S @{}} 
      \toprule
        & \sizehdr{7B} & \sizehdr{$8\times7$B} \\
      \cmidrule(lr){2-4}\cmidrule(lr){5-7}
      \textbf{System} & \metahdr & \metahdr \\
      \midrule
      LLM(Param)    & 0.176 & 0.142 & 0.271 & 0.191 & 0.149 & 0.281 \\
      OrLog(Param)  & 0.122\sigtriangle & 0.112\sigtriangle & 0.221\sigtriangle & 0.163\sigbox & 0.134\sigtriangle & 0.259 \\
      LLM(Param+)   & \bfseries 0.260 & \bfseries 0.180 & \bfseries 0.347 & 0.249 & 0.181 & 0.340 \\
      OrLog(Param+) &  0.160\sigtriangle & 0.139\sigtriangle & .263\sigtriangle & \bfseries 0.266 & \bfseries 0.185\sigbox & \bfseries 0.352 \\
      \bottomrule
    \end{tabular}
  }
\end{subtable}\hfill
\begin{subtable}[t]{0.485\textwidth}
  \centering
  \caption{Qwen-2.5}
  \miniskip
  \resizebox{\linewidth}{!}{%
    \begin{tabular}{@{} l S S S S S S @{}} 
      \toprule
      & \sizehdr{7B} & \sizehdr{72B} \\

      \cmidrule(lr){2-4}\cmidrule(lr){5-7}
      \textbf{System} & \metahdr & \metahdr \\
      \midrule
      LLM(Param)   & 0.181 & 0.146 & 0.273 & 0.204 & 0.160 & 0.298 \\
      OrLog(Param)  & 0.145\sigtriangle & 0.126\sigtriangle & 0.240\sigtriangle & 0.217 & 0.154\sigbullet & 0.304 \\
      LLM(Param+)   & 0.271 & 0.180 & 0.350 & 0.294 & \bfseries 0.201 & \bfseries 0.380 \\
      OrLog(Param+) & \bfseries 0.303\sigbox & \bfseries 0.199\sigtriangle & \bfseries 0.381\sigtriangle & \bfseries 0.300 & 0.196 & 0.374 \\
      \bottomrule
    \end{tabular}
  }
\end{subtable}

\par\smallskip
\miniskip
\begin{subtable}[t]{0.485\textwidth}
  \centering
  \miniskip
  \caption{Llama-3}
  \miniskip
  \resizebox{\linewidth}{!}{%
    \begin{tabular}{@{} l S S S S S S @{}} 
      \toprule
      & \sizehdr{8B} & \sizehdr{70B} \\

      \cmidrule(lr){2-4}\cmidrule(lr){5-7}
      \textbf{System} & \metahdr & \metahdr \\
      \midrule
      LLM(Param)    & .177 & .144  & .269& .192 & .156 & .287 \\
      OrLog(Param)  & .138\sigtriangle & .124\sigtriangle& .234\sigtriangle  & .161\sigbox & .136\sigtriangle & .260\sigtriangle \\
      LLM(Param+)   & .244 & .182 & .339 & .263 & .193 & .355 \\
      OrLog(Param+) & \bfseries .263 & \bfseries .186 & \bfseries .351
                      & \bfseries .308\sigtriangle & \bfseries .197\sigbullet & \bfseries .384\sigbox \\
      \bottomrule
    \end{tabular}
  }
\end{subtable}\hfill

  \label{tab:llm}
\end{table*}

\textit{Negation} shows moderate gains for OrLog. Queries of the form 
\(
A \land \neg B
\)
and 
\(
A \land B \land \neg C
\)
show mean gains in the 0.02--0.03 range, with some structures achieving statistical significance (\(p<0.05\)) with the Qwen backbone. This suggests that negation, while tractable for LLMs, still imposes a mild burden on their end-to-end reasoning, which OrLog alleviates by delegating to ProbLog. The last and most notable is the case of \textit{disjunction}. Simple disjunctions (\(A \lor B\)) exhibit the largest performance gap: OrLog improves NDCG@1 on Llama and on Qwen, both at high significance (\(p<0.05\) and \(p<0.01\), respectively). Even extended disjunctions (\(A \lor B \lor C\)) benefit substantially. These results demonstrate that monolithic LLM reasoning struggles to balance alternative predicate truth‐assignments, whereas OrLog’s symbolic engine effectively integrates multiple probabilistic hypotheses to satisfy at least one of the disjuncts. Qualitative analysis of LLM's results showed that LLMs tend to treat disjunction as conjunction, rejecting a fitting entity for a disjunction by saying that this does not satisfy both conditions.

\vspace*{-2mm}
\section{Discussion and Future Work}

\vspace*{-1mm}
We have proposed OrLog, a framework to resolve complex entity-seeking queries through separation of concerns; decoupling predicate-level plausibility estimation from logical reasoning, using LLMs for the former and probabilistic logic for the latter. We showed empirically that OrLog can achieve significantly better performance than previous approaches using base retrievers or monolithic LLM reasoning. With entity descriptions (Parametric+), OrLog consistently improves top-rank results and shows its largest advantages on disjunctive structures, where the LLM treats disjunctive constraints as conjunction. At the same time, OrLog cuts average tokens by one decoding-free forward pass per predicate and offloading logical reasoning to ProbLog. These results indicate that effective reasoning does not require ever larger models or longer traces; smaller backbones suffice when used as calibrated uncertainty oracles, provided their plausibility outputs are reliable~\cite{belcak2025small}. We have referred to the LLM as an \emph{Oracle}, a deliberate abstraction that assumes access to representative predicate plausibility estimation. However, this assumption is only partially borne out today; the elicited plausibilities from LLMs can be miscalibrated, especially under weak or conflicting evidence. Progress requires improving plausibility elicitation through more sophisticated methods or strengthening the backend model to output more representative scores. To strengthen this component, we see three promising paths for future work; (1) jointly learn plausibility estimation and logical inference via DeepProbLog~\cite{manhaeve2018deepproblog}, a differentiable probabilistic logic programming framework; (2) replace point plausibilities with distributions and marginalize uncertainty during inference; and (3) improve elicitation through better truth valuation prompts and alternate logit-space aggregation. Finally, robustness analysis of OrLog vs LLMs under perturbations of queries or entity descriptions and extending OrLog to other types of complex queries (e.g., multi-hop) constitute the immediate next steps to test whether OrLog's performance gains sustain as noise and task complexity grow. In sum, OrLog demonstrates that principled separation between plausibility estimation and logical reasoning is not only feasible in information retrieval with the tools we have available today, but also a promising foundation for building scalable, reliable reasoning systems.

\subsubsection*{Acknowledgements.}
This research was (partially) supported by the Dutch Research Council (NWO), under project numbers 024.004.022, NWA.1389.20.\-183, and KICH3.LTP.20.006, and the European Union under grant agreements No. 101070212 (FINDHR) and No. 101201510 (UNITE).
All content represents the opinion of the authors, which is not necessarily shared or endorsed by their respective employers and/or sponsors.

\subsubsection*{Disclosure of interests.}
The authors have no competing interests to declare that are relevant to the content of this article. 

\bibliographystyle{splncs04nat}
\bibliography{references} 

@inproceedings{Ng01Advances,
  author =	 {Andrew Y. Ng and Michael I. Jordan and Yair Weiss},
  title =	 {On Spectral Clustering: Analysis and an algorithm},
  booktitle =	 {Advances in Neural Information Processing Systems 14
                  [Neural InformationcProcessing Systems: Natural and
                  Synthetic, {NIPS}]},
  pages =	 {849--856},
  year =	 2001,
}

@article{Lin24ECC,
  author =	 {Zhen Lin and Shubhendu Trivedi and Jimeng Sun},
  title =	 {Generating with Confidence: Uncertainty
                  Quantification for Black-box Large Language Models},
  journal =	 {Trans. Mach. Learn. Res.},
  volume =	 2024,
  year =	 2024,
}

@misc{Kadavath22PE,
  author =	 {Saurav Kadavath and Tom Conerly and Amanda Askell
                  and Tom Henighan and Dawn Drain and Ethan Perez and
                  Nicholas Schiefer and Zac Hatfield{-}Dodds and Nova
                  DasSarma and Eli Tran{-}Johnson and Scott Johnston
                  and Sheer El Showk and Andy Jones and Nelson Elhage
                  and Tristan Hume and Anna Chen and Yuntao Bai and
                  Sam Bowman and Stanislav Fort and Deep Ganguli and
                  Danny Hernandez and Josh Jacobson and Jackson
                  Kernion and Shauna Kravec and Liane Lovitt and Kamal
                  Ndousse and Catherine Olsson and Sam Ringer and
                  Dario Amodei and Tom Brown and Jack Clark and
                  Nicholas Joseph and Ben Mann and Sam McCandlish and
                  Chris Olah and Jared Kaplan},
  title =	 {Language Models (Mostly) Know What They Know},
  year =	 2022,
  eprint =	 {2207.05221},
  archivePrefix ={arXiv},
  primaryClass = {cs.CL},
  url =		 {https://arxiv.org/abs/2207.05221},
}

@inproceedings{Kuhn23SE,
  author =	 {Lorenz Kuhn and Yarin Gal and Sebastian Farquhar},
  title =	 {Semantic Uncertainty: Linguistic Invariances for
                  Uncertainty Estimation in Natural Language
                  Generation},
  booktitle =	 {The Eleventh International Conference on Learning
                  Representations {ICLR}},
  year =	 2023
}

@inproceedings{Bakman24mars,
  author =	 {Yavuz Faruk Bakman and Duygu Nur Yaldiz and Baturalp
                  Buyukates and Chenyang Tao and Dimitrios Dimitriadis
                  and Salman Avestimehr},
  title =	 {{MARS:} Meaning-Aware Response Scoring for
                  Uncertainty Estimation in Generative {LLMs}},
  booktitle =	 {Proceedings of the 62nd Annual Meeting of the
                  Association for Computational Linguistics {ACL}},
  pages =	 {7752--7767},
  year =	 2024
}

@inproceedings{Duan24SAR,
  author =	 {Jinhao Duan and Hao Cheng and Shiqi Wang and Alex
                  Zavalny and Chenan Wang and Renjing Xu and Bhavya
                  Kailkhura and Kaidi Xu},
  title =	 {Shifting Attention to Relevance: Towards the
                  Predictive Uncertainty Quantification of Free-Form
                  Large Language Models},
  booktitle =	 {Proceedings of the 62nd Annual Meeting of the
                  Association for Computational Linguistics (Volume 1:
                  Long Papers), {ACL}},
  pages =	 {5050--5063},
  year =	 {2024}
}

@inproceedings{soudani-etal-2025-uncertainty,
  title =	 "Why Uncertainty Estimation Methods Fall Short in
                  {RAG}: An Axiomatic Analysis",
  author =	 "Soudani, Heydar and Kanoulas, Evangelos and Hasibi,
                  Faegheh",
  booktitle =	 "Findings of the Association for Computational
                  Linguistics: ACL 2025",
  year =	 "2025",
  publisher =	 "Association for Computational Linguistics",
  pages =	 "16596--16616",
}

@inproceedings{DBLP:conf/aaai/ZhangZ0PRRCR25,
  author =	 {Wenhao Zhang and Mengqi Zhang and Shiguang Wu and
                  Jiahuan Pei and Zhaochun Ren and Maarten de Rijke
                  and Zhumin Chen and Pengjie Ren},
  title =	 {ExcluIR: Exclusionary Neural Information Retrieval},
  booktitle =	 {AAAI-25, Sponsored by the Association for the
                  Advancement of Artificial Intelligence},
  year =	 {2025},
}

@inproceedings{DBLP:conf/acl/MalaviyaSCLT23,
  author =	 {Chaitanya Malaviya and Peter Shaw and Ming{-}Wei
                  Chang and Kenton Lee and Kristina Toutanova},
  title =	 {{QUEST:} {A} Retrieval Dataset of Entity-Seeking
                  Queries with Implicit Set Operations},
  booktitle =	 {Proceedings of the 61st Annual Meeting of the
                  Association for Computational Linguistics {ACL}
                  2023},
  publisher =	 {Association for Computational Linguistics},
  year =	 {2023},
}

@inproceedings{weller-etal-2024-nevir,
  author =	 {Orion Weller and Dawn J. Lawrie and Benjamin Van
                  Durme},
  title =	 {NevIR: Negation in Neural Information Retrieval},
  booktitle =	 {Proceedings of the 18th Conference of the European
                  Chapter of the Association for Computational
                  Linguistics, {EACL}},
  publisher =	 {Association for Computational Linguistics},
  year =	 {2024},
}

@inproceedings{van_den_Elsen_2025,
  series =	 {SIGIR ’25},
  title =	 {Reproducing NevIR: Negation in Neural Information
                  Retrieval},
  booktitle =	 {Proceedings of the 48th International ACM SIGIR
                  Conference on Research and Development in
                  Information Retrieval},
  author =	 {van den Elsen, Coen and Barkhof, Francien and
                  Nijdam, Thijmen and Lupart, Simon and Aliannejadi,
                  Mohammad},
  year =	 {2025},
}

@article{10.1145/24634.24635,
  author =	 {van Rijsbergen, C.J.},
  title =	 {A New Theoretical Framework for Information
                  Retrieval},
  year =	 {1986},
  publisher =	 {Association for Computing Machinery},
  journal =	 {SIGIR Forum},
}

@inproceedings{DBLP:conf/nlpir/MaiG024,
  author =	 {Quan Mai and Susan Gauch and Douglas Adams},
  title =	 {SetBERT: Enhancing Retrieval Performance for Boolean
                  Logic and Set Operation Queries},
  booktitle =	 {Proceedings of the 2024 8th International Conference
                  on Natural Language Processing and Information
                  Retrieval},
  year =	 {2024},
}

@inproceedings{DBLP:conf/sigir/BruzaH94,
  author =	 {Peter Bruza and Theo W. C. Huibers},
  title =	 {Investigating Aboutness Axioms using Information
                  Fields},
  booktitle =	 {Proceedings of the 17th Annual International
                  {ACM-SIGIR} Conference on Research and Development
                  in Information Retrieval. (Special Issue of the
                  {SIGIR} Forum)},
  year =	 1994,
}

@article{DBLP:journals/sigir/HuibersLR96,
  author =	 {Theo W. C. Huibers and Mounia Lalmas and C. J. van
                  Rijsbergen},
  title =	 {Information Retrieval and Situation Theory},
  journal =	 {{SIGIR} Forum},
  year =	 {1996},
}

@article{DBLP:journals/jasis/RijsbergenL96,
  author =	 {C. J. van Rijsbergen and Mounia Lalmas},
  title =	 {Information Calculus for Information Retrieval},
  journal =	 {J. Am. Soc. Inf. Sci.},
  year =	 {1996},
}

@Inbook{Crestani1998,
  author =	 "Crestani, Fabio",
  title =	 "Logical Imaging and Probabilistic Information
                  Retrieval",
  bookTitle =	 "Information Retrieval: Uncertainty and Logics:
                  Advanced Models for the Representation and Retrieval
                  of Information",
  year =	 "1998",
  publisher =	 "Springer US",
  address =	 "Boston, MA",
  pages =	 "247--279",
  isbn =	 "978-1-4615-5617-6",
  url =		 "https://doi.org/10.1007/978-1-4615-5617-6\_10"
}

@inproceedings{DBLP:conf/trec/CrestaniRSR95,
  author =	 {Fabio Crestani and Ian Ruthven and Mark Sanderson
                  and C. J. van Rijsbergen},
  title =	 {The Troubles with Using a Logical Model of {IR} on a
                  Large Collection of Documents},
  booktitle =	 {Proceedings of The Fourth Text REtrieval Conference,
                  {TREC} 1995, Gaithersburg, Maryland, USA, November
                  1-3, 1995},
  publisher =	 {National Institute of Standards and Technology
                  {(NIST)}},
  year =	 {1995},
}

@article{DBLP:journals/csur/AbdulahhadBCP19,
  author =	 {Karam Abdulahhad and Catherine Berrut and
                  Jean{-}Pierre Chevallet and Gabriella Pasi},
  title =	 {Modeling Information Retrieval by Formal Logic: {A}
                  Survey},
  journal =	 {{ACM} Comput. Surv.},
  volume =	 52,
  year =	 2019,
}

@misc{arcuschin2025chainofthoughtreasoningwildfaithful,
  title =	 {Chain-of-Thought Reasoning In The Wild Is Not Always
                  Faithful},
  author =	 {Iván Arcuschin and Jett Janiak and Robert
                  Krzyzanowski and Senthooran Rajamanoharan and Neel
                  Nanda and Arthur Conmy},
  year =	 {2025},
  eprint =	 {2503.08679},
  archivePrefix ={arXiv},
}

@article{lanham2023measuring,
  title =	 {Measuring faithfulness in chain-of-thought
                  reasoning},
  author =	 {Lanham, Tamera and Chen, Anna and Radhakrishnan,
                  Ansh and Steiner, Benoit and Denison, Carson and
                  Hernandez, Danny and Li, Dustin and Durmus, Esin and
                  Hubinger, Evan and Kernion, Jackson and others},
  journal =	 {arXiv preprint arXiv:2307.13702},
  year =	 {2023}
}

@article{10.1145/3703155,
  author =	 {Huang, Lei and Yu, Weijiang and Ma, Weitao and
                  Zhong, Weihong and Feng, Zhangyin and Wang, Haotian
                  and Chen, Qianglong and Peng, Weihua and Feng,
                  Xiaocheng and Qin, Bing and Liu, Ting},
  title =	 {A Survey on Hallucination in Large Language Models:
                  Principles, Taxonomy, Challenges, and Open
                  Questions},
  year =	 {2025},
  publisher =	 {Association for Computing Machinery},
  journal =	 {ACM Trans. Inf. Syst.},
}

@inproceedings{
yee2024faithful,
  title =	 {Faithful and Unfaithful Error Recovery in Chain of
                  Thought},
  author =	 {Evelyn Yee and Alice Li and Chenyu Tang and Yeon Ho
                  Jung and Ramamohan Paturi and Leon Bergen},
  booktitle =	 {First Conference on Language Modeling (COLM)},
  year =	 {2024},
}

@techreport{o3o4mini_system_card_2025,
  author =	 {OpenAI},
  title =	 {{OpenAI o3 and o4‑mini System Card}},
  institution =	 {OpenAI},
  year =	 {2025},
  url =		 {https://openai.com/index/o3-o4-mini-system-card/}
}

@article{DBLP:journals/corr/abs-2504-10903,
  author =	 {Sicheng Feng and Gongfan Fang and Xinyin Ma and
                  Xinchao Wang},
  title =	 {Efficient Reasoning Models: {A} Survey},
  journal =	 {CoRR},
  volume =	 {abs/2504.10903},
  year =	 {2025},
}

@inproceedings{DBLP:conf/nips/Wei0SBIXCLZ22,
  author =	 {Jason Wei and Xuezhi Wang and Dale Schuurmans and
                  Maarten Bosma and Brian Ichter and Fei Xia and Ed
                  H. Chi and Quoc V. Le and Denny Zhou},
  title =	 {Chain-of-Thought Prompting Elicits Reasoning in
                  Large Language Models},
  booktitle =	 {Advances in Neural Information Processing Systems
                  35: Annual Conference on Neural Information
                  Processing Systems 2022, NeurIPS 2022, New Orleans,
                  LA, USA, November 28 - December 9, 2022},
  year =	 {2022},
}

@inproceedings{yue2025does,
  title =	 {Does Reinforcement Learning Really Incentivize
                  Reasoning Capacity in {LLM}s Beyond the Base Model?},
  author =	 {Yang Yue and Zhiqi Chen and Rui Lu and Andrew Zhao
                  and Zhaokai Wang and Yang Yue and Shiji Song and Gao
                  Huang},
  booktitle =	 {2nd AI for Math Workshop @ ICML 2025},
  year =	 {2025},
}

@inproceedings{Olausson_2023,
  title =	 {LINC: A Neurosymbolic Approach for Logical Reasoning
                  by Combining Language Models with First-Order Logic
                  Provers},
  booktitle =	 {Proceedings of the 2023 Conference on Empirical
                  Methods in Natural Language Processing},
  publisher =	 {Association for Computational Linguistics},
  author =	 {Olausson, Theo and Gu, Alex and Lipkin, Ben and
                  Zhang, Cedegao and Solar-Lezama, Armando and
                  Tenenbaum, Joshua and Levy, Roger},
  year =	 {2023},
}

@article{yang2023leandojo,
  title =	 {Leandojo: Theorem proving with retrieval-augmented
                  language models},
  author =	 {Yang, Kaiyu and Swope, Aidan and Gu, Alex and
                  Chalamala, Rahul and Song, Peiyang and Yu, Shixing
                  and Godil, Saad and Prenger, Ryan J and Anandkumar,
                  Animashree},
  journal =	 {Advances in Neural Information Processing Systems},
  volume =	 {36},
  pages =	 {21573--21612},
  year =	 {2023}
}

@inproceedings{DBLP:conf/emnlp/PanAWW23,
  author =	 {Liangming Pan and Alon Albalak and Xinyi Wang and
                  William Yang Wang},
  title =	 {Logic-LM: Empowering Large Language Models with
                  Symbolic Solvers for Faithful Logical Reasoning},
  booktitle =	 {Findings of the Association for Computational
                  Linguistics: {EMNLP}},
  publisher =	 {Association for Computational Linguistics},
  year =	 {2023},
}

@inproceedings{DBLP:conf/ijcai/RaedtKT07,
  author       = {Luc De Raedt and
                  Angelika Kimmig and
                  Hannu Toivonen},
  title        = {ProbLog: {A} Probabilistic Prolog and Its Application in Link Discovery},
  booktitle    = {{IJCAI} 2007, Proceedings of the 20th International Joint Conference
                  on Artificial Intelligence},
  year         = {2007},
}

@article{DBLP:journals/corr/abs-2506-02153,
  author =	 {Peter Belcak and Greg Heinrich and Shizhe Diao and
                  Yonggan Fu and Xin Dong and Saurav Muralidharan and
                  Yingyan Celine Lin and Pavlo Molchanov},
  title =	 {Small Language Models are the Future of Agentic
                  {AI}},
  journal =	 {CoRR},
  year =	 {2025},
}

@inproceedings{DBLP:conf/aaai/NafarVK25,
  author =	 {Aliakbar Nafar and Kristen Brent Venable and Parisa
                  Kordjamshidi},
  title =	 {Reasoning over Uncertain Text by Generative Large
                  Language Models},
  booktitle =	 {AAAI-25, Sponsored by the Association for the
                  Advancement of Artificial Intelligence},
  publisher =	 {{AAAI} Press},
  year =	 {2025},
}

@inproceedings{kambhampati2024position,
  title =	 {Position: LLMs can’t plan, but can help planning in
                  LLM-modulo frameworks},
  author =	 {Kambhampati, Subbarao and Valmeekam, Karthik and
                  Guan, Lin and Verma, Mudit and Stechly, Kaya and
                  Bhambri, Siddhant and Saldyt, Lucas Paul and Murthy,
                  Anil B},
  booktitle =	 {Forty-first International Conference on Machine
                  Learning},
  year =	 {2024}
}

@inproceedings{Hoveyda_2025,
  series =	 {SIGIR ’25},
  title =	 {Adaptive Orchestration of Modular Generative
                  Information Access Systems},
  booktitle =	 {Proceedings of the 48th International ACM SIGIR
                  Conference on Research and Development in
                  Information Retrieval},
  publisher =	 {ACM},
  author =	 {Hoveyda, Mohanna and Oosterhuis, Harrie and de
                  Vries, Arjen P. and de Rijke, Maarten and Hasibi,
                  Faegheh},
  year =	 {2025},
}

@article{Wang2022TextEB,
  title =	 {Text Embeddings by Weakly-Supervised Contrastive
                  Pre-training},
  author =	 {Liang Wang and Nan Yang and Xiaolong Huang and
                  Binxing Jiao and Linjun Yang and Daxin Jiang and
                  Rangan Majumder and Furu Wei},
  journal =	 {ArXiv},
  year =	 {2022},
}

@article{DBLP:journals/ftir/RobertsonZ09,
  author =	 {Stephen E. Robertson and Hugo Zaragoza},
  title =	 {The Probabilistic Relevance Framework: {BM25} and
                  Beyond},
  journal =	 {Found. Trends Inf. Retr.},
  volume =	 {3},
  number =	 {4},
  year =	 {2009},
}

@misc{kalai2025languagemodelshallucinate,
  title =	 {Why Language Models Hallucinate},
  author =	 {Adam Tauman Kalai and Ofir Nachum and Santosh
                  S. Vempala and Edwin Zhang},
  year =	 {2025},
  eprint =	 {2509.04664},
  archivePrefix ={arXiv},
  primaryClass = {cs.CL},
}

@article{belcak2025small,
  title =	 {Small Language Models are the Future of Agentic AI},
  author =	 {Belcak, Peter and Heinrich, Greg and Diao, Shizhe
                  and Fu, Yonggan and Dong, Xin and Muralidharan,
                  Saurav and Lin, Yingyan Celine and Molchanov, Pavlo},
  journal =	 {arXiv preprint arXiv:2506.02153},
  year =	 {2025}
}

@article{manhaeve2018deepproblog,
  title =	 {Deepproblog: Neural probabilistic logic programming},
  author =	 {Manhaeve, Robin and Dumancic, Sebastijan and Kimmig,
                  Angelika and Demeester, Thomas and De Raedt, Luc},
  journal =	 {Advances in neural information processing systems},
  volume =	 {31},
  year =	 {2018}
}

@misc{weller2025theoreticallimitationsembeddingbasedretrieval,
  title =	 {On the Theoretical Limitations of Embedding-Based
                  Retrieval},
  author =	 {Orion Weller and Michael Boratko and Iftekhar Naim
                  and Jinhyuk Lee},
  year =	 {2025},
  eprint =	 {2508.21038},
  archivePrefix ={arXiv},
  primaryClass = {cs.IR},
}

@article{petcu2025comprehensive,
  title =	 {A Comprehensive Taxonomy of Negation for {NLP} and
                  Neural Retrievers},
  author =	 {Petcu, Roxana and Bhargav, Samarth and de Rijke,
                  Maarten and Kanoulas, Evangelos},
  journal =	 {arXiv preprint arXiv:2507.22337},
  year =	 {2025}
}

@misc{deepmind2024imo,
  author =	 {DeepMind},
  title =	 {AI Achieves Silver-Medal Standard Solving
                  International Mathematical Olympiad Problems},
  howpublished =
                  {\url{https://deepmind.google/discover/blog/ai-solves-imo-problems-at-silver-medal-level/}},
  note =	 {Accessed: 2025-10-01},
  year =	 {2024},
  month =	 {July},
  day =		 {25}
}

@article{DBLP:journals/corr/abs-2412-05563,
  author =	 {Ola Shorinwa and Zhiting Mei and Justin Lidard and
                  Allen Z. Ren and Anirudha Majumdar},
  title =	 {A Survey on Uncertainty Quantification of Large
                  Language Models: Taxonomy, Open Research Challenges,
                  and Future Directions},
  journal =	 {CoRR},
  volume =	 {abs/2412.05563},
  year =	 {2024},
  doi =		 {10.48550/ARXIV.2412.05563},
}

@article{DBLP:journals/corr/abs-2507-12547,
  author =	 {Lionel Wong and Katherine M. Collins and Lance Ying
                  and Cedegao E. Zhang and Adrian Weller and Tobias
                  Gerstenberg and Timothy O'Donnell and Alexander
                  K. Lew and Jacob Andreas and Joshua B. Tenenbaum and
                  Tyler Brooke{-}Wilson},
  title =	 {Modeling Open-World Cognition as On-Demand Synthesis
                  of Probabilistic Models},
  journal =	 {CoRR},
  volume =	 {abs/2507.12547},
  year =	 {2025},
}

@misc{wong2023wordmodelsworldmodels,
  title =	 {From Word Models to World Models: Translating from
                  Natural Language to the Probabilistic Language of
                  Thought},
  author =	 {Lionel Wong and Gabriel Grand and Alexander K. Lew
                  and Noah D. Goodman and Vikash K. Mansinghka and
                  Jacob Andreas and Joshua B. Tenenbaum},
  year =	 {2023},
  eprint =	 {2306.12672},
  archivePrefix ={arXiv},
  primaryClass = {cs.CL},
}

@book{Polya1968-PLYMAP,
  address =	 {Princeton, N.J.,},
  author =	 {George Po?lya},
  editor =	 {},
  publisher =	 {Princeton University Press},
  title =	 {Mathematics and Plausible Reasoning},
  year =	 {1968}
}

@book{Jaynes2002-JAYPTT,
  author =	 {Edwin T. Jaynes},
  publisher =	 {Cambridge University Press: Cambridge},
  title =	 {Probability Theory. The Logic of Science},
  year =	 {2002}
}

@inproceedings{zhuang-etal-2024-beyond,
    title = "Beyond Yes and No: Improving Zero-Shot {LLM} Rankers via Scoring Fine-Grained Relevance Labels",
    author = "Zhuang, Honglei  and
      Qin, Zhen  and
      Hui, Kai  and
      Wu, Junru  and
      Yan, Le  and
      Wang, Xuanhui  and
      Bendersky, Michael",
    booktitle = "Proceedings of the 2024 Conference of the North American Chapter of the Association for Computational Linguistics: Human Language Technologies (Volume 2: Short Papers)",
    year = "2024",
    publisher = "Association for Computational Linguistics",
}

@inproceedings{nogueira-etal-2020-document,
    title = "Document Ranking with a Pretrained Sequence-to-Sequence Model",
    author = "Nogueira, Rodrigo  and
      Jiang, Zhiying  and
      Pradeep, Ronak  and
      Lin, Jimmy",
    booktitle = "Findings of the Association for Computational Linguistics: EMNLP 2020",
    year = "2020",
    publisher = "Association for Computational Linguistics",
}

@inproceedings{shen-etal-2025-logicol,
    title = "{L}ogi{C}o{L}: Logically-Informed Contrastive Learning for Set-based Dense Retrieval",
    author = "Shen, Yanzhen  and
      Chen, Sihao  and
      Xu, Xueqiang  and
      Zhang, Yunyi  and
      Malaviya, Chaitanya  and
      Roth, Dan",
    booktitle = "Proceedings of the 2025 Conference on Empirical Methods in Natural Language Processing",
    year = "2025",
    publisher = "Association for Computational Linguistics",
}
\end{document}